# Analysis of the inverse square-root size effect in the plasticity of metals

T.T. Zhu[a], B. Ehrler[b], X.D. Hou[a], A.J. Bushby[a,*]

and D.J. Dunstan[b]

Centre for Materials Research,

[a] School of Engineering and Materials Science,

[b] Department of Physics,

Queen Mary, University of London, Mile End, E1 4NS, UK.

*Corresponding Author

Dr Andy Bushby
School of Engineering and Materials Science
Queen Mary, University of London
Mile End Road
London E1 4NS, UK

Tel +44 20 7882 5276
Fax + 44 20 8981 9804
E-mail a.j.bushby@qmul.ac.uk

**Abstract**

Small-scale mechanical behaviour shows significant departures from classical elastic-plastic theory. In a remarkable number of instances, the strength of a material appears to scale as the reciprocal square root of the smallest length scale. There are several recent experimental and modeling results in the literature that show an interaction between dimensional (extrinsic) size and microstructural (intrinsic) size effects. In this paper, we present a mechanical model that naturally produces the inverse square root strengthening and derive an expression for the 'effective length' when both the extrinsic and intrinsic size effects are significant. The theory fits well to data from a wide range of deformation geometries and includes the interaction between the microstructural and dimensional size effects. Furthermore, this approach is able to predict the size effect under uniform deformation without strain gradient.

# 1. Introduction

Small-scale mechanical behaviour is at the cutting-edge of research in materials science and applied mechanics. It has been known for several decades that materials display a strong size effect, in which the strength is enhanced when the size of the structure or of the stressed volume is decreased. For metals such as gold, copper or nickel, this occurs when the characteristic length associated with deformation is less than the order of tens of microns. Generally, the origin of the size effect can be divided into microstructural (intrinsic) and dimensional (extrinsic) characteristic lengths. Microstructural size effects include those due to grain boundaries (Hall, 1951; Petch, 1953) and particle reinforcement (Lloyd, 1994). Effects of dimensional size have been presented for compression of pillars (Greer et al., 2005; Volkert and Lilleodden, 2006) torsion of wires (Fleck et al., 1994), bending of foils (Stölken and Evans, 1998; Moreau et al., 2005; Ehrler et al., 2008), indentation (Ma and Clarke, 1995; Nix and Gao, 1998; Lim and Chaudhri 1999; Swadener et al., 2002; Spary et al., 2006; Zhu et al., 2008a, c), and other geometries (Nix, 1989; Espinosa et al., 2004). In a remarkable number of instances material strength scales as the inverse square root of the smallest length scale (specimen dimension, grain size, indentation contact radius and so on) (Volkert and Lilleodden, 2006; Hall, 1951; Petch, 1953; Ma and Clarke, 1995; Zhu et al., 2008a, c): herein designated as $1/\sqrt{l}$ scaling. In the majority of these size dependent cases, strain gradients are involved, and there is general agreement that the size effect can then be attributed to hardening due to geometrically-necessary dislocations (GNDs) (Ashby, 1970). A characteristic length $l^*$ is often introduced to parameterise the theory. However, in other cases, strain gradients are not involved; for instance, micro-pillar compression has uniform deformation without strain gradients, but also appears to closely follow $1/\sqrt{l}$ scaling (Volkert and Lilleodden, 2006). General models capable of predicting $1/\sqrt{l}$ for such a diverse range of tests and microstructures have yet to emerge and it is not apparent why such an exact relationship should hold. Strain gradient plasticity theories (Fleck and Hutchinson, 1993; Nix and Gao, 1998; Huang et al., 2000a, b; Han et al., 2005) can predict $1/\sqrt{l}$ scaling for some situations. However, there are various forms of the theory that depend on how the contributions from statistically stored and geometrically necessary dislocations are formulated. The physical interpretation of the characteristic length $l^*$ in these theories is not clear and the values of $l^*$ can be inconsistent. For instance, in indentation of Cu a value of $l^* \approx 30\mu m$ is obtained (Huang et al., 2000b) while in the twisting of Cu wires $l^* \approx 4\mu m$ (Huang et al., 2000a). Similarly there are a number of explanations for the Hall-Petch type grain size dependent strengthening, often associated with dislocation pile-up at grain boundaries (Zhu et al., 2008b), while other approaches consider dislocation density (Conrad et al., 1967) and dislocation source density (Widjaja et al, 2007a; Motz, et al., 2008).

Interaction between dimensional (extrinsic) and microstructure (intrinsic) size effects has been found in the strengthening of thin films on a substrate (Venkatraman and Bravman, 1992; Keller et al., 1998; Espinosa et al., 2004). Keller et al. (1998) and Espinosa et al. (2004) argued that the Hall-Petch effect and thickness effect should be additive. However, Thompson (1992) proposed that a dislocation storage mechanism occurs in thin layers at both the interface with the substrate and at grain boundaries.

Thompson's theory leads to a symmetrical dependence of strength on the inverse (rather than the inverse square root) of grain size and film thickness. Recently, Cao et al. (2006) studied the indentation size effect on polycrystalline Ag and Au films on Si substrates and found that the Hall-Petch behaviour is additive to the indentation size effect. Ehrler et al. (2008) and Hou et al. (2008) have recently reported precise experimental results on the interaction between dimensional (extrinsic) effects and grain size from foil bending and from spherical indentation respectively which suggest a more complex interaction. Discrete dislocation dynamics simulations have also investigated the interaction of indentation size effect and grain size (Widjaja et al., 2007a) and found similar relationships to the experiments of Hou et al. (2008).  An understanding of the combination of size effects is essentially necessary for applied mechanics and engineering. Existing theories for combining microstructure and extrinsic size effects are rarely found in the literature.

No existing theory predicts $1/\sqrt{l}$ behaviour both for the situations where $l$ is the order of the grain size, $d$, and where $l$ is the order of the structure size, $h$.  Still less are there any theoretical approaches which address the situation where $l$ is the order of both $d$ and $h$, when $h$ is not the characteristic length in strain gradient theory.  Our approach is to reconsider the classic theory of Conrad et al. (1967) in which the key factor is the distance that a dislocation can move.  This dislocation mean free path has recently been identified as the key quantity in the strength and hardening behaviour of soft metals (Devincre et al., 2008) and cited as a governing behaviour in polycrystalline materials and in size effects associated with small volumes.  This theory readily captures the size effect with and without strain gradient, and in any case without requiring the characteristic length $l^*$ used in strain gradient plasticity theory.  Then, in a new analysis of the interactions of grain size and structure size, we consider what the effective length scale should be when $d$ is the order of $h$, thus defining the length scale parameter $l_{eff}$.  This approach gives a model in excellent agreement with a wide variety of experimental data spanning ranges of geometry, structure size and grain size, and including cases with and without a strain gradient.

## 2.  Models for size effects

First of all, we introduce an analysis which essentially attributes the $1/\sqrt{l}$ behaviour to dislocation density (Taylor hardening (1934)) under fairly standard assumptions but without requiring a strain gradient.

### 2.1. Mechanical analysis of slip distance

Under homogeneous loading, mobile dislocations are considered to account for the plastic strain. Each of them is supposed to travel a mean free path, an average distance $\bar{x}$, limited by obstacles such as grain boundaries, or indeed by encountering a free surface. The plastic strain is:

$$\varepsilon_{pl} = \rho_m b \bar{x} \tag{1}$$

where $\rho_m$ is the mobile dislocation density and $b$ is the effective Burgers vector. We take the mean free path $\bar{x}$ to be proportional to the characteristic length scale $l$, such as the distance to a grain boundary or a free surface,

$$\bar{x} = l / \lambda \tag{2}$$

where $\lambda$ is a proportionality coefficient of the order of unity. Let the mobile dislocation density be proportional to the total dislocation density,

$$\rho_m = \xi\rho \tag{3}$$

where $\xi$ is a proportionality coefficient and with $0 < \xi < 1$. Then the plastic strain is:

$$\varepsilon_{pl} = (\xi/\lambda)\rho bl \tag{4}$$

Consider an elementary event in which a dislocation is forced past a pair of pinning points at a separation $r$, illustrated in Fig.1. The maximum length of the part of the dislocation between the points is ½π$r$ (the semicircle) and afterwards this part of the dislocation snaps back to length $r$ (the diameter). So the energy dissipated as heat is (½π − 1) $E_d$, where $E_d$ is the dislocation self-energy per unit length (J/m) and is generally written as $E_d = \alpha\mu b^2$, where $\mu$ is the shear modulus and $\alpha$ is a constant of the order of unity. We identify the pinning points as other dislocations, so that $r$ is the average separation between dislocations is, $r = 1/\sqrt{\rho}$, and for a dislocation to move, an elementary event has to happen on average once for every $r$ of dislocation line length and for every $r$ of distance moved.

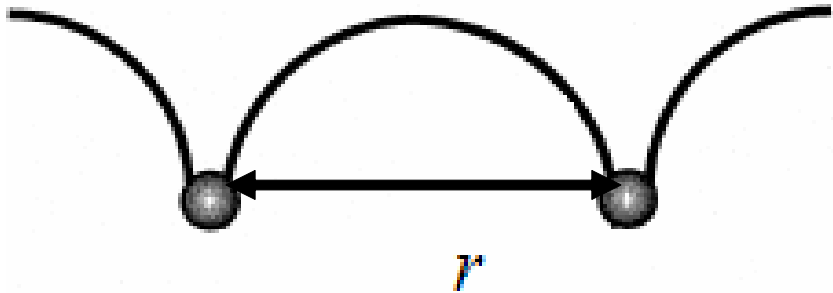

$r$

Fig. 1. Schematic illustration of a dislocation forced past a pair of pinning points at a separation $r$.

We take as an *Ansatz* that at all times the plastic strain is given by Eq. (1). The total line length of the mobile dislocations is $L$ in the strained volume. Then an increment of plastic strain implies an increment of dislocation density,

$$d\varepsilon_{pl} = \bar{x} b d\rho_m \tag{5}$$

and an increment in the total line length of dislocations,

$$dL = V d\rho_m = l^3 d\rho_m = l^3 \xi d\rho \tag{6}$$

These extra dislocations have all moved the distance $\bar{x}$ given by Eq. (2). The number of elementary events is then

$$dN_{events} = \frac{dL}{r}\frac{\bar{x}}{r} = \frac{\frac{1}{\lambda}l dL}{r^2} = \frac{\xi}{\lambda}\frac{l^4 d\rho}{r^2} = \frac{\xi}{\lambda} l^4 \rho d\rho = \frac{l^3 \rho}{b} d\varepsilon_{pl} = \frac{\lambda}{\xi}\frac{l^2}{b^2}\varepsilon_{pl} d\varepsilon_{pl} \tag{7}$$

The plastic energy per unit volume dissipated as heat is then (approximating ½π − 1 to 1)

$$\begin{aligned}dW &= \frac{dN_{events}}{l^3} r E_d = \frac{\lambda}{\xi}\frac{1}{lb^2} r E_d \varepsilon_{pl} d\varepsilon_{pl} \\ &= \frac{\lambda}{\xi}\frac{1}{lb^2}\frac{1}{\sqrt{\rho}} E_d \varepsilon_{pl} d\varepsilon_{pl} = \left(\frac{\lambda}{\xi}\right)^{1/2}\frac{1}{lb^2}\frac{\sqrt{lb}}{\sqrt{\varepsilon_{pl}}} E_d \varepsilon_{pl} d\varepsilon_{pl}\end{aligned} \tag{8}$$

or,

$$dW = \left(\frac{\lambda}{\xi}\right)^{1/2}\sqrt{\frac{\varepsilon_{pl}}{lb^3}} E_d d\varepsilon_{pl} \tag{9}$$

With $E_d = \alpha \mu b^2$, and with an initial yield stress $\tau_0$, the flow stress is:

$$\tau = \tau_0 + \frac{dW}{d\varepsilon_{pl}} = \tau_0 + \alpha\left(\frac{\lambda}{\xi}\right)^{1/2}\sqrt{\frac{\varepsilon_{pl}}{lb^3}}\mu b^2 = \tau_0 + \alpha\left(\frac{\lambda}{\xi}\right)^{1/2}\mu\sqrt{b}\frac{\sqrt{\varepsilon_{pl}}}{\sqrt{l}} \tag{10}$$

Note that $1/\sqrt{l}$ strengthening is predicted, and it is also linked directly with square-root power-law work-hardening.

*2.2. Model for slip distance analysis*

The derivation in section 2.1 is essential that of the Taylor forest hardening mechanism in that the primary interaction occurs between dislocation and that their spacing is proportional to dislocation density.   This theory relies on the idea that dislocation sources are available within the material and these generate dislocations that interact with other dislocations within the material's volume. Evidence from recent dislocation dynamics simulations (Weygand et al., 2008; Benzerga and Shaver, 2006) suggest that even in the small volumes of material associated with micro-pillar compression experiments, there are sufficient internal sources and interactions to support this idea.   Experimentally, several slip systems are often observed to operate, even in crystals oriented for single slip (Volkert and Lilleodden, 2006; Weygand et al., 2008).    Hence we might expect such a mechanism to operate at length scales of a few hundred nanometres and above. At small length scales other mechanisms might prevail, such as surface nucleation in pillars or grain boundary nucleation and grain boundary sliding in the bulk (Benzerga and Shaver, 2006; Van Swygenhoven, 2002).   In the slip distance analysis the size effect arises independently of strain gradient. The density of geometrically necessary dislocations is anyway proportional to the plastic strain and so is subsumed into the coefficient $\xi$ of Eqs.(3) and (4). This coefficient thus subsumes the characteristic length $l^*$ used in strain gradient plasticity theory.

The tensile flow stress is related to the shear flow stress by,

$$\sigma = A\tau \qquad (11)$$

where *A* is the Taylor factor, which may be interpreted as an isotropic expression of the crystalline anisotropy at the continuum level. A value of *A* = 3.06 is given for fcc metals (Gao et al., 1999, Kocks, 1970). Then,

$$\sigma = \sigma_0 + AC\mu\sqrt{b}\,\frac{\sqrt{\varepsilon_{pl}}}{\sqrt{l}} \qquad (12)$$

where *C* is the coefficient $C = \alpha(\lambda/\xi)^{1/2}$ of the order of unity.

## 3. Characteristic length for polycrystalline materials

In single crystal structures, the characteristic length *l* which determines the effective slip distance $\bar{x}$, is relatively easy to identify. For instance, in pillar compression, it is the diameter *D* of the pillar; in bending, it is the thickness *h* of the foil; in twisting, it is the diameter of the wires and for nanoindentation, we expect it to be the contact radius *a* of the indent (Zhu et al., 2008c). However, it will be more complicated in polycrystalline materials, because of the influence of the grain size, *d*.

We expect the extrinsic size effect and the intrinsic size effect due to grain size to interact when they are both of significant scale. The mean slip distance should be determined by an effective length $l_{eff}$ determined both by grain size and by dimensional (extrinsic) constraints. In what follows, we shall suppose that the grain structure is independent of the dimensional constraint – as if, say, a foil was cut out of a bulk material after the grain structure was formed.

Consider a one-dimensional external size constraint, for instance, a foil undergoing bending, with thickness *h* and grain size *d*, as shown in Figs. 2 and 3. A dislocation may ideally be taken to pass along the *h* direction, as indicated in the figure by the broken arrow, although, in reality, its slip plane in different grains would be at various angles to the *h* direction.

Noted that in the limiting case with $d > h$, the dislocation slip distance $\bar{x}$ is only influenced and scaled by *h*, because the dislocation will not generally encounter grain boundaries but can only experience the boundaries delimiting *h* (the free surfaces). On the other hand, in the other limit, $d < h$, $\bar{x}$ can be simply taken as scaled by *d*, because a moving dislocation generally encounters grain boundaries before it encounters the foil surface.

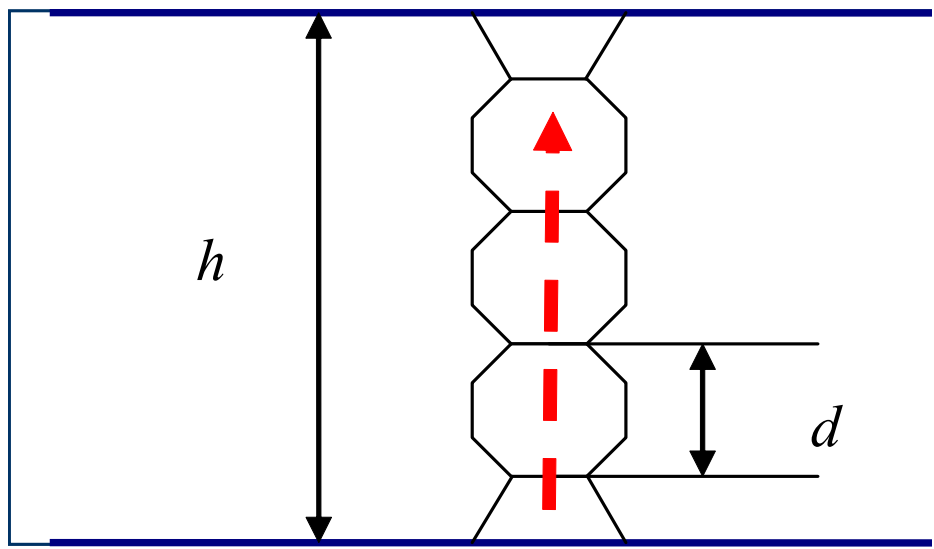

(a)

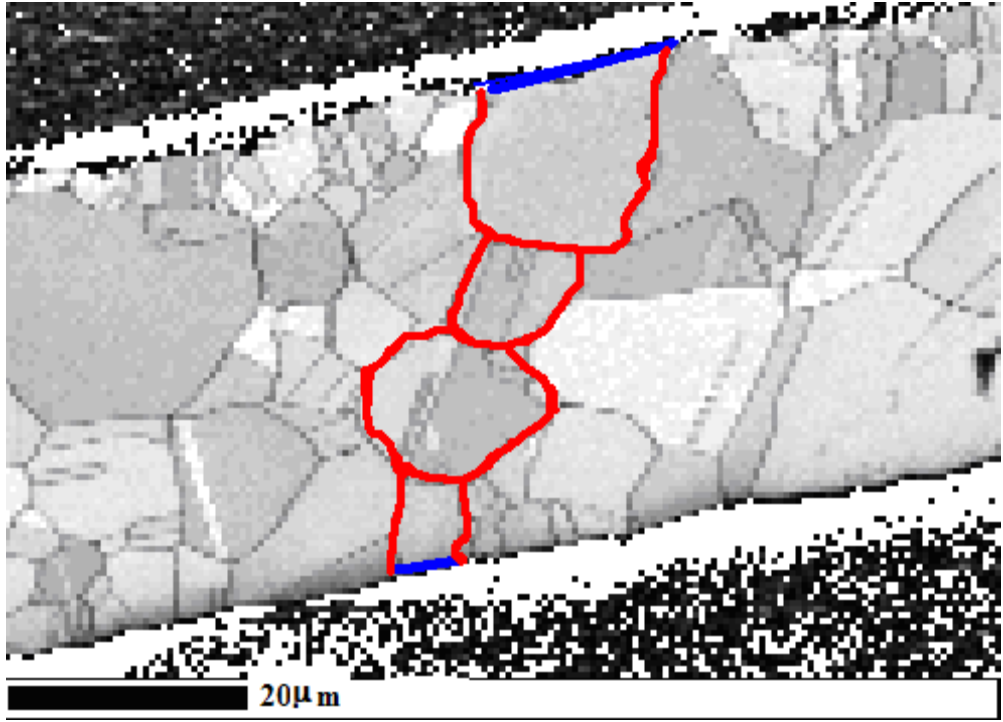

(b)

Fig. 2. (a) Schematic diagram of a foil cross section of thickness of $h$ with grain size $d$. The broken arrow shows a path available to dislocations along the thickness direction.   (b) Electron backscattering diffraction (EBSD) orientation map of the cross-section of a 50μm nickel foil with the average grain size $d = 14$μm. Some grain boundaries are highlighted to show a structure resembling the schematic of (a).

Fig.2 is drawn to illustrate the case where $d$ and $h$ are comparable but $d < h$.   If the grain structure is independent of the presence of free surfaces, then we have part-grains at both surfaces of foil (here drawn as half-grains). A path length $h$ traversing $N$ grains would imply a value for $l_{eff}$ of

$$l_{eff} = \frac{h}{N} \qquad (13)$$

However, the number of internal whole grains may be counted as $\frac{h}{d} - 1$. Counting the two surface part-grains as grains, we add 2, giving

$$N = \frac{h}{d} + 1 \qquad (14)$$

Substituting Eq. (14) into Eq. (13), we obtain:

$$l_{eff} = \frac{h}{N} = \frac{h}{\frac{h}{d} + 1} = \frac{1}{\frac{1}{h} + \frac{1}{d}} \qquad (15)$$

which can also be written as

$$\frac{1}{l_{eff}} = \frac{1}{d} + \frac{1}{h} \qquad (16)$$

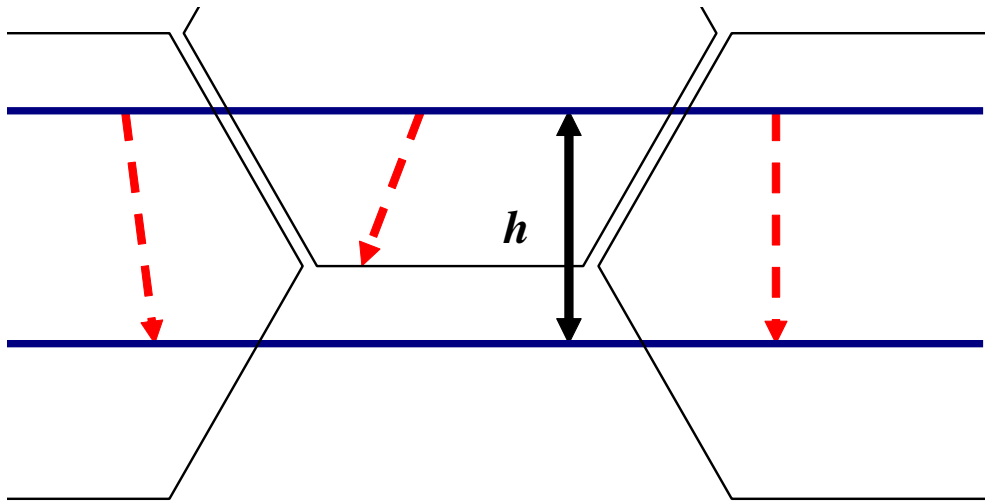

(a)

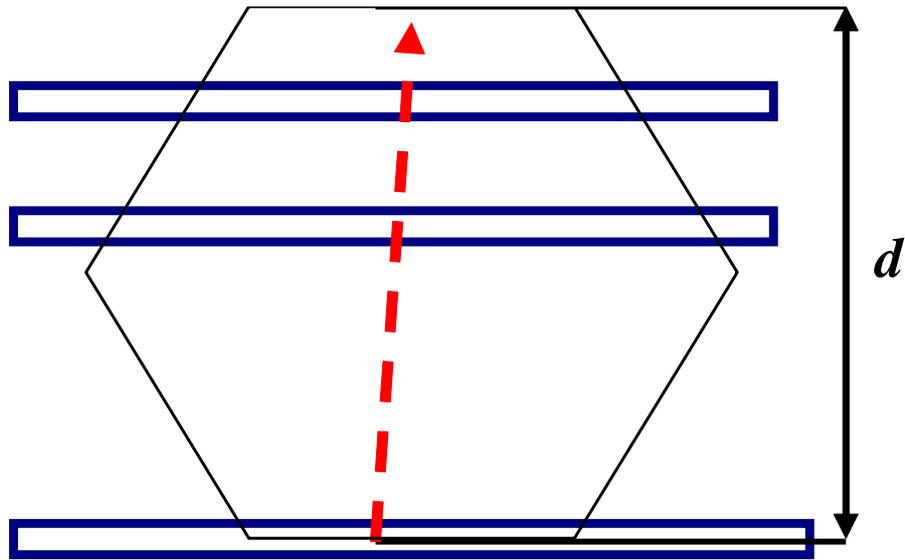

(b)

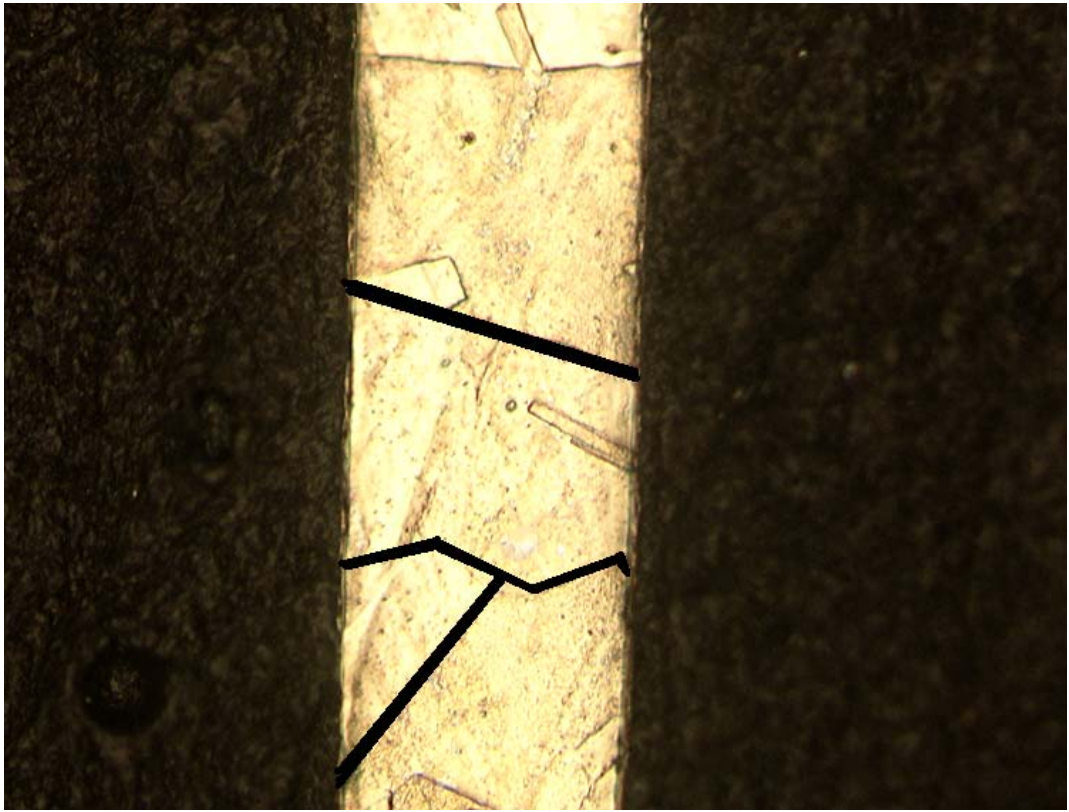

(c)

Fig.3. (a): Schematic diagram of a foil with thickness $h$ and grain size $d$, for $d > h$. The broken arrows indicate paths available to dislocations. Some dislocation path lengths are equal to $h$, while others are stopped at internal grain boundaries. In (b), an alternative way

of considering this situation is shown. A foil section may be cut out of a large-grain material at a random position relative to the grain shown. In most cases, the whole foil thickness is available to provide a path to a dislocation. However, *h*/*d* foils contain a grain boundary, so that the available path is only part of the foil thickness. In (c), an optical microscopic image of a cross section of a125μm nickel foil with an average grain size of 200μm is shown. Some grains extend from one free surface to the other, while elsewhere grain boundaries appear within the thickness, as in the schematic diagram of (a).

We must also consider the situation where *d* and *h* are comparable but *d* > *h*, as shown in Fig. 3(a), in which again the paths of dislocations are shown by broken arrows. Some may have the whole distance *h* available from surface to surface in a single grain, while others will be influenced by a grain boundary.

In Fig. 3(b), this situation is represented as a path length through a whole grain, length *d*, but interrupted by free surface boundaries. This figure is directly analogous to Fig. 2(a), and the analysis is the same, except that Eq. (13) becomes

$$l_{eff} = \frac{d}{N} \tag{17}$$

Now we have

$$N = \frac{d}{h} + 1 \tag{18}$$

Substituting Eq. (18) into Eq. (17), we obtain

$$l_{eff} = \frac{d}{N} = \frac{d}{\frac{d}{h}+1} = \frac{1}{\frac{1}{h}+\frac{1}{d}} \tag{19}$$

which again can be written as

$$\frac{1}{l_{eff}} = \frac{1}{d} + \frac{1}{h} \qquad (20)$$

So we have the same expression, Eqs. (16) and (20), for $l_{eff}$ in the cases $d < h$ and $d > h$. This expression also has the correct behaviour in the limits $d \gg h$ and $d \ll h$. Note, however, that it is obtained assuming that the dislocations slip in the direction of $h$. Other directions have greater lengths (with a $1/\cos\theta$ term), and so in fits to experimental data we may expect a non-unity coefficient on $h^{-1}$.

Experimental measurements may check this model. As an example, a cross-section electron backscatter diffraction (EBSD) image (EBSD, HKL5, Oxford Instruments, UK) of a nickel foil is shown in Fig. 2(b) The sample possess 50μm thickness and with grain size around 14μm, i.e. $d < h$. In Fig. 2(b), the grain boundaries are highlighted to illustrate a structure resembling the schematic of Fig. 2(a). Fig. 3(c) shows an optical microscopic cross section picture of another nickel foil, with 125μm thickness and with grain size around 200μm, i.e. $d > h$. Some grains extend from one free surface to the other, while elsewhere grain boundaries appear within the thickness, as in the schematic diagram of Fig. 3(a).

In the case of nanoindentation, the plastic zone extends radially in three dimensions. Under an axi-symmetric indenter (conical, spherical, and including as an approximation Berkovich and cube corner indenters), the contact radius is $a$. Following Johnson (1985), we consider the effective deformation zone as a hemisphere which scales with contact size $a$ and hence volume $V = \frac{2}{3}\pi a^3$, shown schematically in Fig. 4. A three-dimensional analogue of the foregoing argument would consider the number of part-grains (grains intersecting the hemispherical surface), which is of order $2\pi a^2 / d^2$, and conclude that the number of grains participating in the deformation is therefore not $V/d^3$ but $V/d^3 + 2\pi a^2 / d^2$. (See the schematic diagram shown in Fig. 4). As above, an effective length, $l_{eff}$, can then be derived.

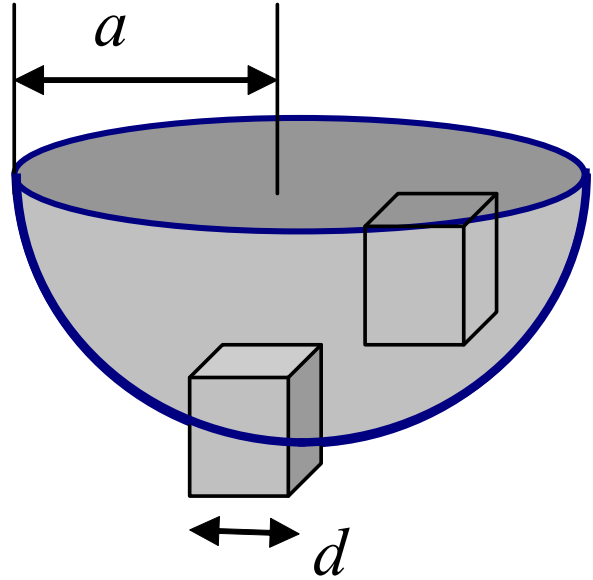

Fig. 4. Schematic of the plastic deformation zone beneath a spherical indentation with contact radius $a$. The plastic zone is approximated to a hemisphere of radius $a$, after Johnson (1987). Grains in the metal are considered as cubic in shape with side length $d$, some of which may be wholly within the plastic zone and others partially within.

More simply, we note that what matters are only the slip planes and the distances dislocations can move along them. Thus the one-dimensional analysis above should remain applicable. However, the volume is not a sphere and has a different size in $z$ ($a$) and in $x$ and $y$ ($2a$). So we expect different effective lengths in these directions,

$$\frac{1}{l_x} = \frac{1}{l_y} = \frac{1}{d} + \frac{1}{2a} \tag{21}$$

$$\frac{1}{l_z} = \frac{1}{d} + \frac{1}{a} \tag{22}$$

Some suitable average should be taken, e.g.,

$$\frac{1}{l_{eff}} = \frac{1}{3}\left(\frac{1}{l_x} + \frac{1}{l_y} + \frac{1}{l_z}\right) = \frac{1}{d} + \frac{2}{3a} \tag{23}$$

Thus, as in the one-dimensional case above, we may expect fits to experiment to require a non-unity coefficient on $a^{-1}$.

## 4. Comparison with experiment

Now we apply this model to several sets of experimental data from the literature, covering a range of geometries and grain sizes. We begin with the uniaxial stress state for both single crystal and polycrystalline samples, then consider the flexure of thin foils and indentation of polycrystalline metals.

*4.1 Uniaxial stress*

The uniaxial compression testing of micro-pillars has been reported recently by several groups (Greer et al., 2004; Volkert and Lilleodden, 2006; Uchic et al., 2004). These experiments present a strong size effect in the absence of any significant strain gradients. The experimental data we use here is taken from Volkert and Lilleodden (2006) for single-crystal gold pillar compression (orientated for single slip). They plotted the compression stress at 5% strain against the pillar diameter, $D$, in a logarithmic plot and found that the best fit was $\sigma \propto D^{-0.61}$. Here, we replot the data against $D^{-1/2}$ (Fig.5). The dotted line is a fit for the stress calculated from slip distance theory, Eq. (12), with $l = D$,

$$\sigma = \sigma_0 + AC\mu\sqrt{b}\frac{\sqrt{\varepsilon_{pl}}}{\sqrt{D}} \tag{24}$$

and with the numerical values $A = 3.06$ (for fcc metals (Gao et al., 1999; Kocks, 1970), $\sigma_0 \approx 0$, $\mu = 28$GPa, $b = 0.25$nm and $\varepsilon_{pl} = 0.05$. $C$ is treated as the only free fitting parameter. The best fit is obtained for $C = 0.52 \pm 0.01$. Scatter in the data is seen to increase as $D^{-1/2}$ increases (decreasing pillar diameter), since the probability of pop-in (elastic overload) increases associated with activation of specific dislocation sources.

To check whether the value of $C$ is reasonable, we recall that $C = \alpha \left( \lambda / \xi \right)^{1/2}$. Values of $\lambda$ and $\xi$ can be estimated. $\lambda$ is a coefficient to characterize the proportion of dislocation mean free path length to the characteristic size. In pillar deformation, considering dislocation slip at 45º, the dislocation can travel more than the characteristic length $D$. A reasonable value for $\lambda$ would be $1/\sqrt{2}$. Considering the proportionality coefficient, $\xi$, (fraction of mobile dislocations in the total dislocation density) it has been proposed that dislocations are not retained inside the pillar, new dislocations are continually nucleated and immediately glide to the free surface, (Greer et al., 2004; Benzera and Shaver, 2006; Uchic et al., 2004; Shan et al., 2008). Correspondingly, in this model, the proportion of mobile dislocations might be relatively high and implies that $\xi = 1$. Then, $C = 0.48$ gives $\alpha = 0.57$.

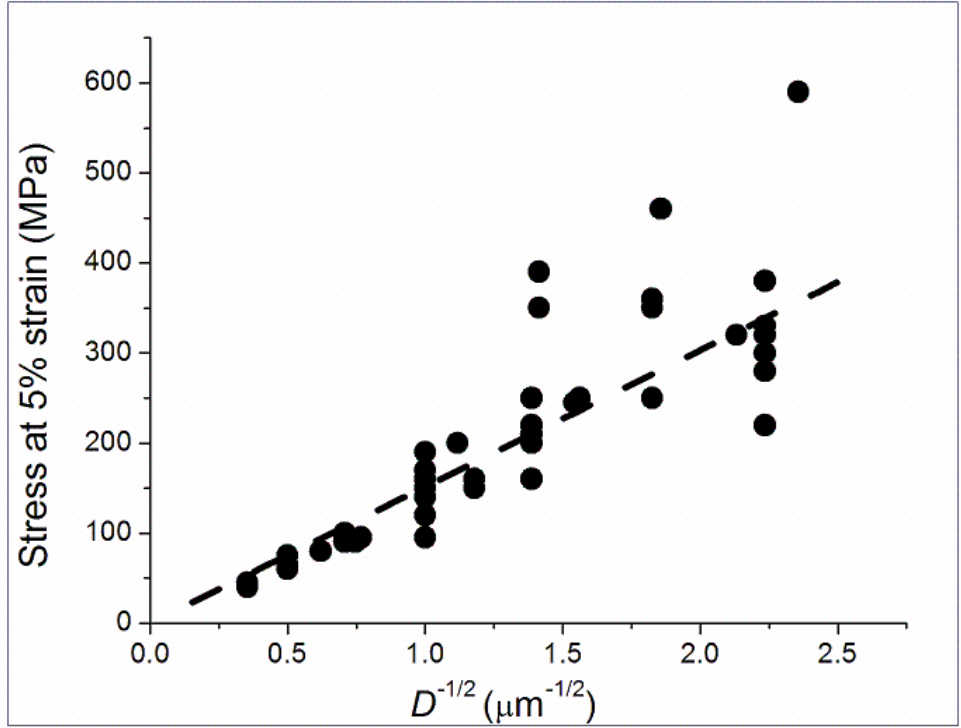

Fig. 5. Single-crystal gold pillar compression data replotted from Volkert and Lilleodden (2006). The stress at 5% strain is plotted against the inverse square-root of the pillar diameter, $D$. The dashed line is the theoretical prediction using Eq. (27), with the best fit value of the parameter, $C = 0.52 \pm 0.01$ ($R^2 = 0.729$).

Another example of uniaxial loading in very thin polycrystalline metal foils comes from bulge testing experiments. Micro bulge test of thin films reported recently (Xiang et al., 2005; Vlassak and Nix, 1992; Xiang and Vlassak, 2006) has the advantage of precise sample fabrication and minimal sample handling. Since the deflection of the film is far smaller than the film width, the test can be considered as plain tension, i.e, without a strain gradient (Vlassak and Nix, 1992). Xiang and Vlassak (2006) carried out bulge test on polycrystalline thin copper films with different surface conditions, i.e., passivated and unpassivated. Here, we only use the data for unpassivated films.

Fig. 6 replots the data for the film stress at $\varepsilon_{pl}$ = 0.02%. The dotted line is a fit for the stress calculated from slip distance theory, Eq. (12), with effective length $l_{eff}$ in Eq. (16):

$$\sigma = \sigma_0 + AC\mu\sqrt{b}\sqrt{\frac{1}{d}+\frac{1}{h}}\sqrt{\varepsilon_{pl}} \qquad (25)$$

and with the numerical values $A$ = 3.06 (for fcc metals (Gao et al., 1999; Kocks et al., 1970)), $\sigma_0 \approx$ 45MPa, $\mu$ = 43.5GPa, $b$ = 0.25nm and $\varepsilon_{pl}$ = 0.02. $h$ is the film thickness and $d$ is the grain size of the foil. $C$ is treated as the only free fitting parameter. The best fit is obtained for $C$ = 1.34 ± 0.02.

Now, we consider whether the value of $C$ is reasonable, where we need to apply again that $C$ is a function of $\lambda$, $\xi$ and $\alpha$, as $C = \alpha\,(\lambda/\xi)^{1/2}$. $\lambda$ and $\xi$ can be estimated. We apply here: $\lambda$ = 1/√2. The fraction of mobile dislocation density could be taken as $\xi \approx$ 0.3 from Hackelöer et al. (1977). In this case, $C$ = 1.34 gives $\alpha$ = 0.87.

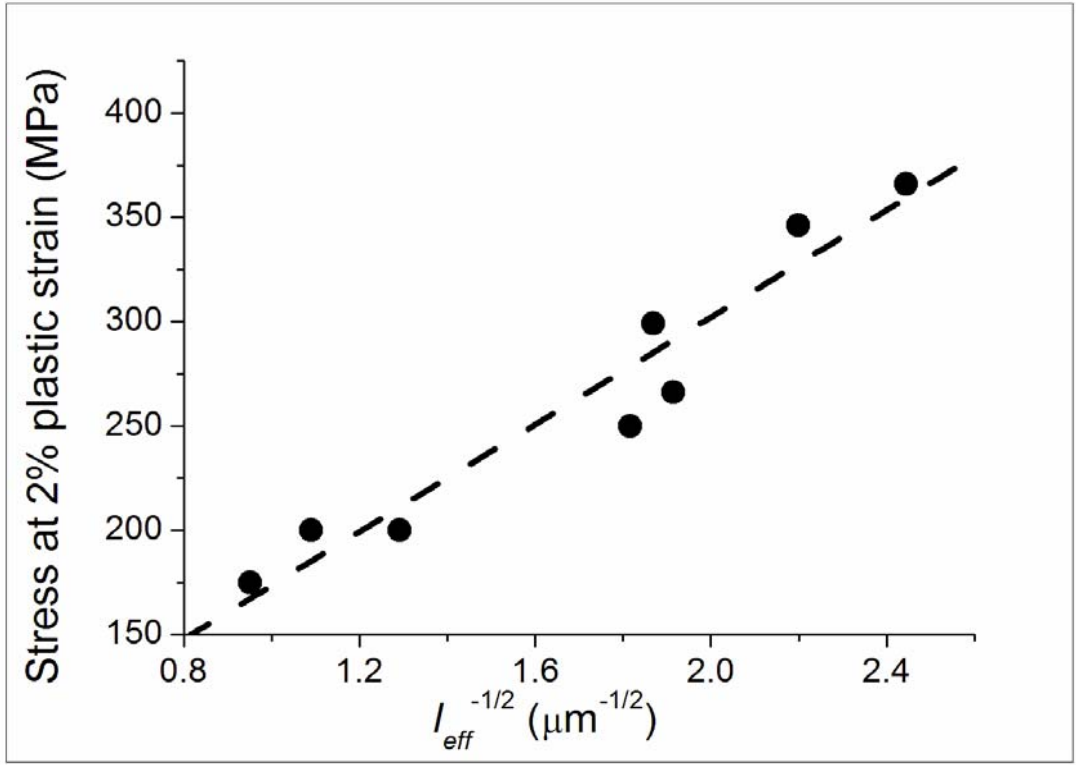

Fig. 6. Copper bulge test data replotted from Xiang and Vlassak (2006). The stress at 0.2% plastic strain is plotted versus $l_{eff}^{-1/2}$ (where $\frac{1}{l_{eff}} = \frac{1}{d} + \frac{1}{h}$). The dashed line is the theoretical prediction using Eq. (25), with the best fit value of the parameter, $C = 1.34 \pm 0.02$ ($R^2 = 0.931$)

*4.2 Bending size effect*

We have recently made studies of the bending moment induced in nickel foils as a function of curvature (Moreau et al., 2005). The full data-set and details of the experimental methods is published elsewhere (Ehrler et al., 2008). The experiments are based on a design implemented by Stölken and Evans (1998) in which foils were bent over mandrels of known radius ('load') and the then released ('unload'). The 'load' curvature gives the strain and the surface strain is defined as $\varepsilon_s = h\kappa/2$, where $\kappa$ is the load curvature. On the other hand, the reduction in curvature in the 'unload' state gives the bending moment, or stress. With technical improvements to the experiment, we have obtained much more precise data over a much wider range of strain, and we have also varied grain size by rapid thermal annealing of the foils before testing. Three foils with different thickness of 10μm, 50μm and 125μm were tested, with grain sizes ranging from 6μm to 200μm. Ratios of $d/h$ range from 3 to 0.03, thus providing a stringent test of expressions in $d^{-1} + h^{-1}$.

In order to compare the theory with obtained experimental results, we evaluate the normalized bending moment $M_n$. It is conveniently normalized by the widths $w$ and thicknesses $h$ of the different foils. Then,

$$M_n = \frac{2\int_0^{\frac{h}{2}} \sigma(z) z \, dz}{wh^2} \qquad (26)$$

Fig. 6 shows the experimental data for normalized bending moment at $\varepsilon_s = 0.1\%$. To calculate the theoretical normalized bending moment, we apply the slip distance expression Eq. (12) for the stress $\sigma(z)$, and for the effective length scale $l_{eff}$ in polycrystalline bending we use Eq. (20), where $h$ is taken to be the foil thickness. Then,

$$\sigma = \sigma_0 + AC\mu\sqrt{b}\sqrt{\frac{1}{d} + \frac{1}{h}}\sqrt{\varepsilon_{pl}} \qquad (27)$$

We approximate that $\sigma_0 \approx 0$, and so throughout the foil, $\varepsilon_{pl} = z\kappa$. Then,

$$\sigma(z) = AC\mu\sqrt{b}\sqrt{\frac{1}{d} + \frac{1}{h}}\sqrt{z\kappa} \qquad (28)$$

Integrating Eq. (30), we obtain the normalized bending moment as,

$$M_n = \frac{1}{5} AC\mu\sqrt{b}\sqrt{\frac{1}{d}+\frac{1}{h}}\sqrt{\varepsilon_s} \qquad (29)$$

The theoretical fit is obtained from Eq. (28), with parameter values of $\varepsilon_s = 0.1\%$, $A = 3.06$ (for fcc metals (Gao et al., 1999; Kocks, 1970)), $\sigma_0 \approx 0$, $\mu = 78$GPa and $b = 0.245$nm and with $C$ as the only fitting variable. A best fit is obtained by applying $C = 2.8 \pm 0.05$. The result is plotted as the dotted line on Fig. 6. It can be seen that the fitted strength with foil thickness and grain size is highly consistent with the measurements.

Now, we consider whether the value of $C$ is reasonable, where we need to apply again that $C$ is a function of $\lambda$, $\xi$ and $\alpha$, as $C = \alpha (\lambda / \xi)^{1/2}$. $\lambda$ and $\xi$ can be estimated. For foil bending, since dislocations would not traverse the neutral plane, the mean free path of the dislocations is smaller than half of the characteristic length. We apply here: $\lambda=3$. The fraction of mobile dislocation density could be taken as $\xi \approx 0.3$ from Hackelöer et al. (1977). In this case, $C = 2.8$ gives $\alpha = 0.88$.

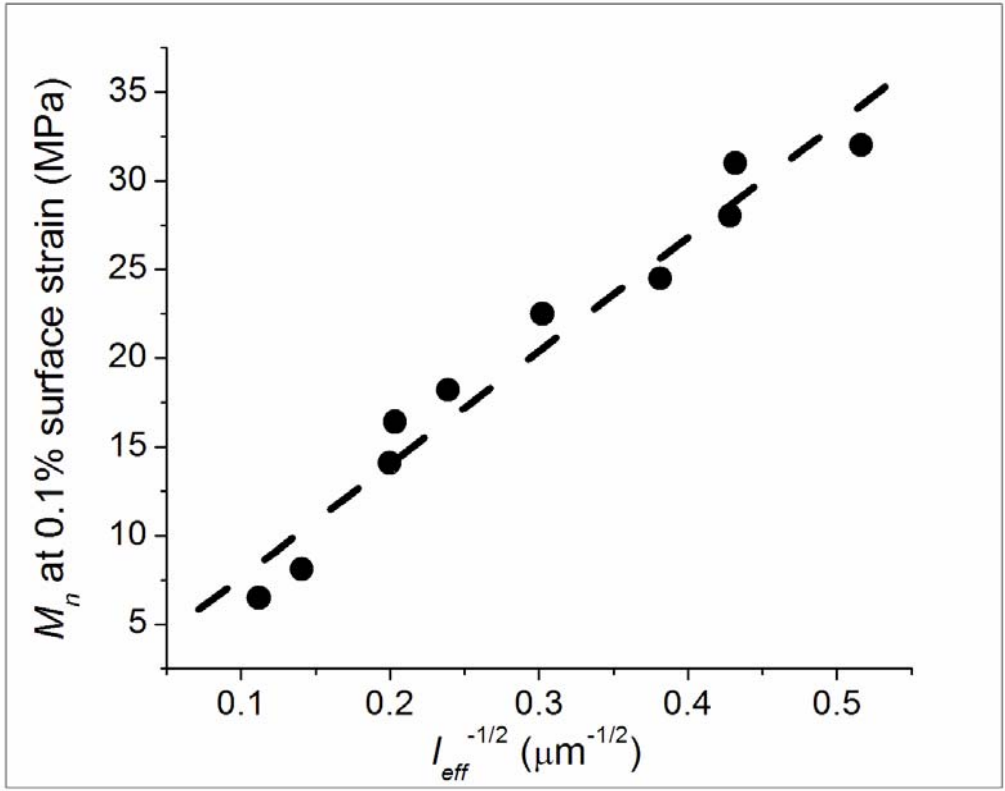

Fig. 7. Nickel foil bending data replotted from Erhler et al. (2008) as normalized bending moment $M_n$ at 0.1% surface strain versus $l_{eff}^{-1/2}$ (where $\frac{1}{l_{eff}} = \frac{1}{d} + \frac{1}{h}$). The dashed line is the theoretical prediction using Eq. (31), with the best fit value of the parameter, $C = 2.8 \pm 0.05$ ($R^2 = 0.951$).

*4.3 Indentation size effect*

Recently, Hou et al. (2008) performed spherical nanoindentation on single crystal and polycrystalline copper. They used different radius indenters on samples with a range of grain sizes. Their data shows interaction between the grain size effect and the indentation size effect. Three indenters were used with the contact radii *a* ranging from 0.82μm to 50μm. The grain sizes varied from 1.15μm to infinite large (single crystal), giving ratios of *d / a* in the range from infinite to 0.3. This data set is therefore able to provide a stringent test of slip distance theory and of size effects that use expressions in $d^{-1} + a^{-1}$. The results for indentation mean pressure measured at an indentation strain of 0.25 are replotted from Hou et al. (2008) in Fig.7.

In order to compare these results with the theory of Section 2, we evaluate the indentation mean pressure $P_m$ as (Tabor, 1951),

$$P_m = 2.8\sigma \tag{30}$$

To calculate the theoretical indentation mean pressure, we use again the slip distance theory expression Eq. (12) for stress $\sigma$, and for the effective length scale $l_{eff}$ in polycrystalline indentation we use Eq. (23). Then,

$$\sigma = \sigma_0 + AC\mu\sqrt{b}\sqrt{\frac{1}{d} + \frac{2}{3a}}\sqrt{\varepsilon_{pl}} \tag{31}$$

Again, we approximate that $\sigma_0 \approx 0$. Under these assumptions, it is possible to evaluate the plastic strain in indentation (Johnson, 1985) as

$$\varepsilon_{pl} = 0.2\varepsilon_{ind} = 0.2\frac{a}{R} \tag{32}$$

where $\varepsilon_{ind}$ is the indentation strain, conventionally defined as *a/R* for spherical indentation (Johnson, 1985). The theoretical prediction of $P_m$ is therefore,

$$P_m = 2.8\left(AC\mu\sqrt{b}\sqrt{\frac{1}{d} + \frac{2}{3a}}\sqrt{0.2\varepsilon_{ind}}\right) \tag{33}$$

Using Eq. (32), with parameter values of $\varepsilon_{ind}$ = 0.25, *A* = 3.06 (for fcc metals (Gao, 1999; Kocks, 1970)), $\sigma_0 \approx 0$, $\mu$ = 43.5GPa, *b* = 0.256nm, *C* is the only free fitting parameter. A best fit is obtained by using *C* = 0.84 ± 0.02, plotted as the dotted line in Fig 7.

The predicted variation of strength with contact radius and grain size is consistent with the measurements.

Now, we consider whether the value of $C$ is reasonable. We recall that $C = \alpha (\lambda/\xi)^{1/2}$. In indentation, a dislocation is supposed to move across the plastic zone. Since the indentation contact size $a$ is the characteristic length here while the plastic zone radius is about $3a$ (Hou et al., 2008), a dislocation is supposed to travel farther than the characteristic length. It is reasonable to take $\lambda \approx 1/3$. As for $\xi$, from Hackelöer et al. (1970), the fraction of mobile dislocations could be taken as $\xi \approx 0.3$. Then, $C = 0.84$ gives $\alpha = 0.80$.

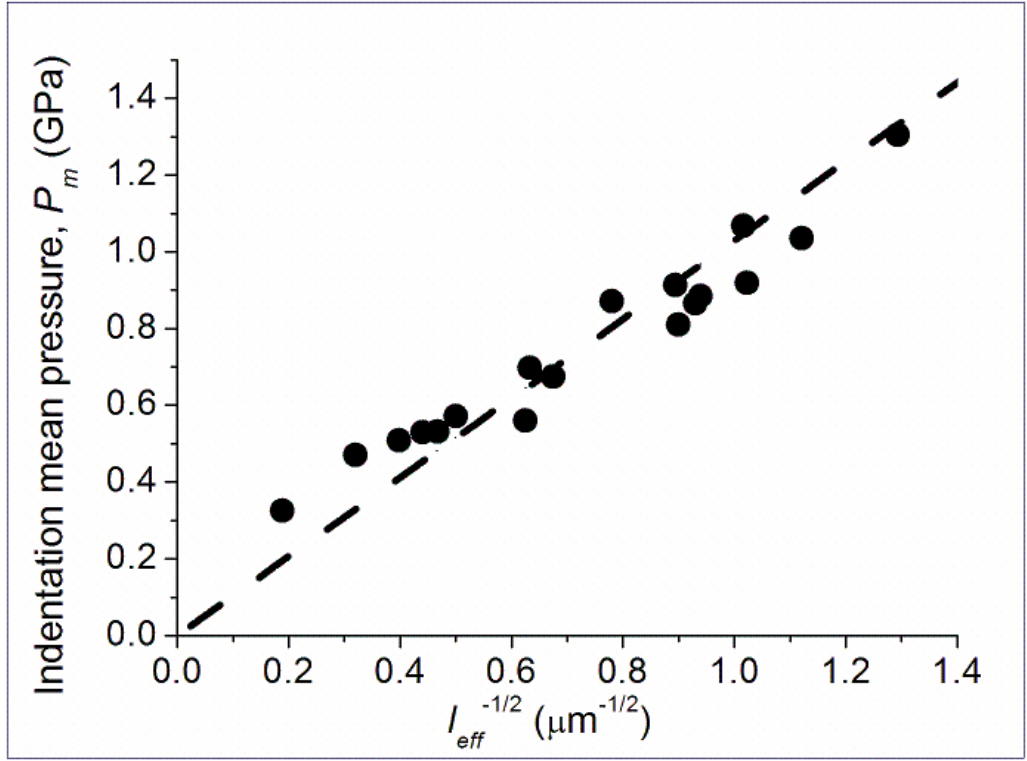

Fig. 8. Polycrystalline copper indentation experimental data replotted from Hou et al. (2008) as $P_m$ (at 0.25 indentation strain) against $l_{eff}^{-1/2}$ (where $\dfrac{1}{l_{eff}} = \dfrac{1}{d} + \dfrac{2}{3a}$). The dashed line is the theoretical prediction using Eq. (35), with the best fit value of the parameter, $C = 0.84 \pm 0.02$ ($R^2 = 0.803$).

Remarkably, the wedge indentation size effect via two-dimensional discrete dislocation plasticity, carried out by Widjaja et al. (2007a) recently, agrees with this theory as well. In these data, due to the finite size of the block in their simulation, the elastic hardness is *h*-dependent (Widjaja et al., 2007b). So, in Fig. 9, the plotted hardness is normalised by the elastic hardness (Widjaja et al., 2007a).

The theoretical pressure is obtained by using Eq. (33), with parameter values of $\varepsilon_{ind}$ = 0.01, $A$ = 3.06 (for fcc metals (Gao et al., 1999; Kocks, 1970), $\sigma_0 \approx 0$, $\mu$ = 26.3GPa, $b$ = 0.286nm, $C$ is the only free fitting parameter. A best fit is obtained by using $C$ = 1.01 ± 0.02, plotted as the dotted line in Fig. 8.7. The predicted variation of strength with contact radius and grain size is very consistent with the measurements. With inputting the same parameter $\lambda$ and $\xi$ as for polycrystalline copper indentation ($\lambda$ = 1/3 and $\xi \approx 0.3$), then $C$ = 1.01 gives $\alpha$ = 0.95.

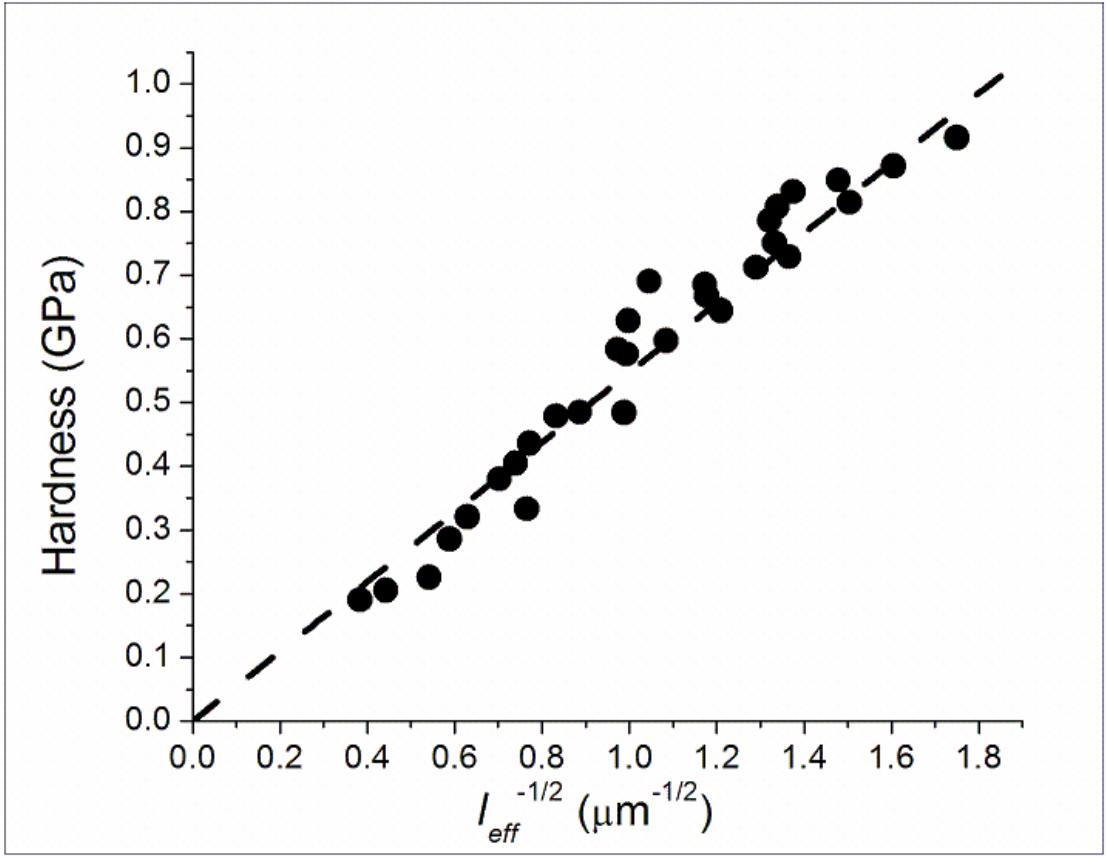

Fig. 9 Simulated wedge indentation data for polycrystalline material from dislocation dynamics simulations of Widjaja *et al.* (2007a) plotted as nominal hardness against $l_{eff}^{-1/2}$ (where $\frac{1}{l_{eff}} = \frac{1}{d} + \frac{2}{3a}$). The dashed line is the theoretical prediction using Eq. (33), with the best fit value of the parameter $C = 1.01 \pm 0.02$ ($R^2 = 0.957$).

## 5. Discussion

Recent dislocation dynamics simulations (Devincere et al., 2008; Weygand et al., 2007b; Benzerga and Shaver, 2006) suggest that the dislocation source size, the spacing of pinning points and hence source arm length, or the mean free path, dictate the initial yield and low plastic strain work-hardening behaviour in metals. The slip distance model predicts the inverse square root size effect of plasticity in metals in general deformation with or without a strain gradient and, in either case, it is without an extra characteristic length parameter $l^*$. Furthermore, this model has also been shown to predict the indentation size effect for both metals and ceramics (Bushby et al., 2009).

The only free fitting parameter $C$ in the theory is obtained from the best fit for experimental data. The value of $C$ is found by this process to an accuracy of about ±2%. In all the different loading geometries, by estimating the value of the coefficients $\lambda$ (the proportion of the dislocation mean free path to the characteristic size) and $\xi$ (fraction of mobile dislocations), the coefficient $\alpha$ is obtained. The value of $\alpha$ is always in a narrow range, which is between 0.57 and 0.88. This is in a good agreement with the cited values in the literature (between 0.2 and 1.2 (Gao et al., 1999; Huang et al., 2004)). The value of $C$ is seen to be mostly dominated by $\lambda$, the coefficient of proportionality between the material length scale and the distance a dislocation can move. In the case of a micro-pillar this is limited by the structure size while in the case of indentation it may be several times the contact radius. So the values obtained for $C$ would appear to be intuitive and could be predicted from sensible estimates of $\alpha$, $\lambda$ and $\tilde{\xi}$

The ability of the approach presented here to combine different length scales, both intrinsic and extrinsic thus defining an effective length scale, is compelling. In the bending of thin foils the grain boundaries and free surfaces appear to have the same effect in limiting the slip distance and controlling the size effect. Similarly in the case of indentation the grain boundaries and extent of plastic zone also appear to present the same limitation of dislocation movement. These results imply that increased strength at small length scales can be achieved however the slip distance is delineated; by grain boundaries, free surfaces or strain gradients.

## 6. Conclusions

Comparison with experiments in a diverse range of loading geometries, both in uniform and non-uniform deformation, shows that the theory of 'slip distance' is consistent with the experimental observations showing $1/\sqrt{l}$ scaling. Within the theory, different intrinsic and extrinsic length scales can be successfully combined to define an effective length scale for the material, $l_{eff}$, as has been suggested previously in the context of strain gradient plasticity theory but without requiring a strain gradient.

## Acknowledgements

The authors thank Prof. A.G. Evans (University of California, USA) and Dr. N.M. Jennett (National Physical Laboratory, UK) for useful discussions, and acknowledge EPSRC for financial support (under grant # EP/C518004/1).

Figure captions

Fig. 1. Schematic illustration of a dislocation forced past a pair of pinning points at a separation *r*.

Fig. 2. (a) Schematic diagram of a foil cross section of thickness of *h* with grain size *d*. The broken arrow shows a path available to dislocations along the thickness direction.   (b) Electron backscattering diffraction (EBSD) orientation map of the cross-section of a 50μm nickel foil with the average grain size *d* = 14μm. Some grain boundaries are highlighted to show a structure resembling the schematic of (a).

Fig.3. (a): Schematic diagram of a foil with thickness *h* and grain size *d*, for *d* > *h*. The broken arrows indicate paths available to dislocations. Some dislocation path lengths are equal to *h*, while others are stopped at internal grain boundaries. In (b), an alternative way of considering this situation is shown. A foil section may be cut out of a large-grain material at a random position relative to the grain shown. In most cases, the whole foil thickness is available to provide a path to a dislocation. However, *h*/*d* foils contain a grain boundary, so that the available path is only part of the foil thickness. In (c), an optical microscopic image of a cross section of a125μm nickel foil with an average grain size of 200μm is shown. Some grains extend from one free surface to the other, while elsewhere grain boundaries appear within the thickness, as in the schematic diagram of (a).

Fig. 4. Schematic of the plastic deformation zone beneath a spherical indentation with contact radius *a*. The plastic zone is approximated to a hemisphere of radius *a*, after Johnson (1987).   Grains in the metal are considered as cubic in shape with side length *d*, some of which may be wholly within the plastic zone and others partially within.

Fig. 5.   Single-crystal gold pillar compression data replotted from Volkert and Lilleodden (2006). The stress at 5% strain is plotted against the inverse square-root of the pillar diameter, *D*. The dashed line is the theoretical prediction using Eq. (27), with the best fit value of the parameter, $C = 0.52 \pm 0.01$ ($R^2 = 0.729$).

Fig. 6. Copper bulge test data replotted from Xiang and Vlassak (2006). The stress at 0.2% plastic strain is plotted versus $l_{eff}^{-1/2}$ (where $\frac{1}{l_{eff}} = \frac{1}{d} + \frac{1}{h}$). The dashed line is the theoretical prediction using Eq. (25), with the best fit value of the parameter, $C = 1.34 \pm 0.02$ ($R^2 = 0.931$).

Fig. 7. Nickel foil bending data replotted from Erhler et al. (2008) as normalized bending moment $M_n$ at 0.1% surface strain versus $l_{eff}^{-1/2}$ (where $\frac{1}{l_{eff}} = \frac{1}{d} + \frac{1}{h}$). The dashed line is the theoretical prediction using Eq. (31), with the best fit value of the parameter, $C = 2.8 \pm 0.05$ ($R^2 = 0.951$).

Fig. 8. Polycrystalline copper indentation experimental data replotted from Hou et al. (2008) as $P_m$ (at 0.25 indentation strain) against $l_{eff}^{-1/2}$ (where $\frac{1}{l_{eff}} = \frac{1}{d} + \frac{2}{3a}$). The dashed line is the theoretical prediction using Eq. (35), with the best fit value of the parameter, $C = 0.84 \pm 0.02$ ($R^2 = 0.803$).

Fig. 9 Simulated wedge indentation data for polycrystalline material from dislocation dynamics simulations of Widjaja *et al.* (2007a) plotted as nominal hardness against $l_{eff}^{-1/2}$ (where $\frac{1}{l_{eff}} = \frac{1}{d} + \frac{2}{3a}$). The dashed line is the theoretical prediction using Eq. (33), with the best fit value of the parameter $C = 1.01 \pm 0.02$ ($R^2 = 0.957$).